\documentclass[12pt]{iopart}

\usepackage{graphicx}
\usepackage{dcolumn}
\usepackage{caption}

\newcommand {\roots}    {\ensuremath{\sqrt{s}}}

\renewcommand {\pt}       {\ensuremath{p_{T} }}  
\newcommand {\AJ}       {\ensuremath{A_J}}

\newcommand {\PbPb}  {\mbox{PbPb}}

\begin{document}

\begin{figure}[t!]
\begin{center}
\includegraphics[width=1.\textwidth]{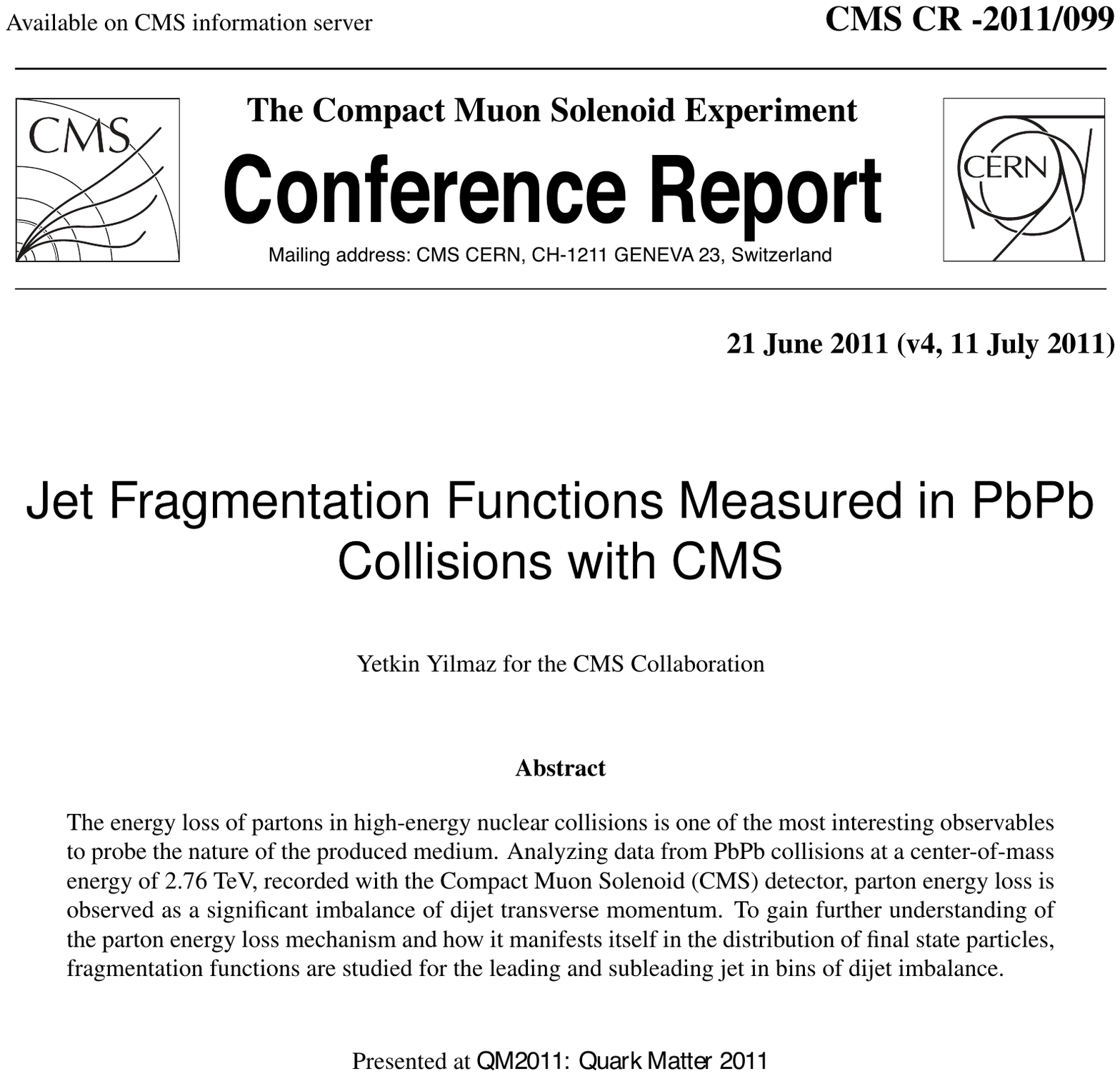}
\end{center}
\end{figure}

\title[Jet Fragmentation Functions Measured in PbPb Collisions with CMS]{Jet Fragmentation Functions Measured in PbPb Collisions with CMS}

\author{Yetkin Yilmaz for the CMS Collaboration}

\address{Massachusetts Institute of Technology, 77 Massachusetts Ave. Cambridge MA 02139-4307, USA}
\ead{yetkin.yilmaz@cern.ch}
\begin{abstract}

The energy loss of partons in high-energy nuclear collisions is one of the most interesting observables to probe the nature of the produced medium. Analyzing data from PbPb collisions at a center-of-mass energy of 2.76 TeV, recorded with the Compact Muon Solenoid (CMS) detector, parton energy loss is observed as a significant imbalance of dijet transverse momentum. To gain further understanding of the parton energy loss mechanism and how it manifests itself in the distribution of final state particles, fragmentation functions are studied for the leading and subleading jet in bins of dijet imbalance. 

\end{abstract}


One of the experimental signatures of Quark-Gluon-Plasma (QGP) formation is the attenuation or disappearance of the spray of hadrons
resulting from the fragmentation of a hard scattered parton having suffered energy loss
in the QGP i.e.,``jet quenching''~\cite{Bjorken:1982tu}. Recent results on PbPb collisions from CMS~\cite{Chatrchyan:2011sx} and ATLAS~\cite{Collaboration:2010bu} experiments have shown that dijets become increasingly imbalanced with event centrality; and the CMS results have highlighted that most of the quenched energy is transferred out of the jet cone which suggests that the reconstructed jet represents the fragments of the parton after quenching. In this study~\cite{FragmentationPAS} we investigate to what extent the fragmentation pattern of the remnant parton resembles vacuum fragmentation.

The fragmentation functions are constructed from charged tracks in the jet cone. We restrict ourselves to the high \pt\ component of the fragmentation function, using charged tracks of \pt\ $ > $ 4 GeV/c, where background contribution is negligible.

\begin{figure}[b!]
\begin{center}
\includegraphics[width=1.0\textwidth]{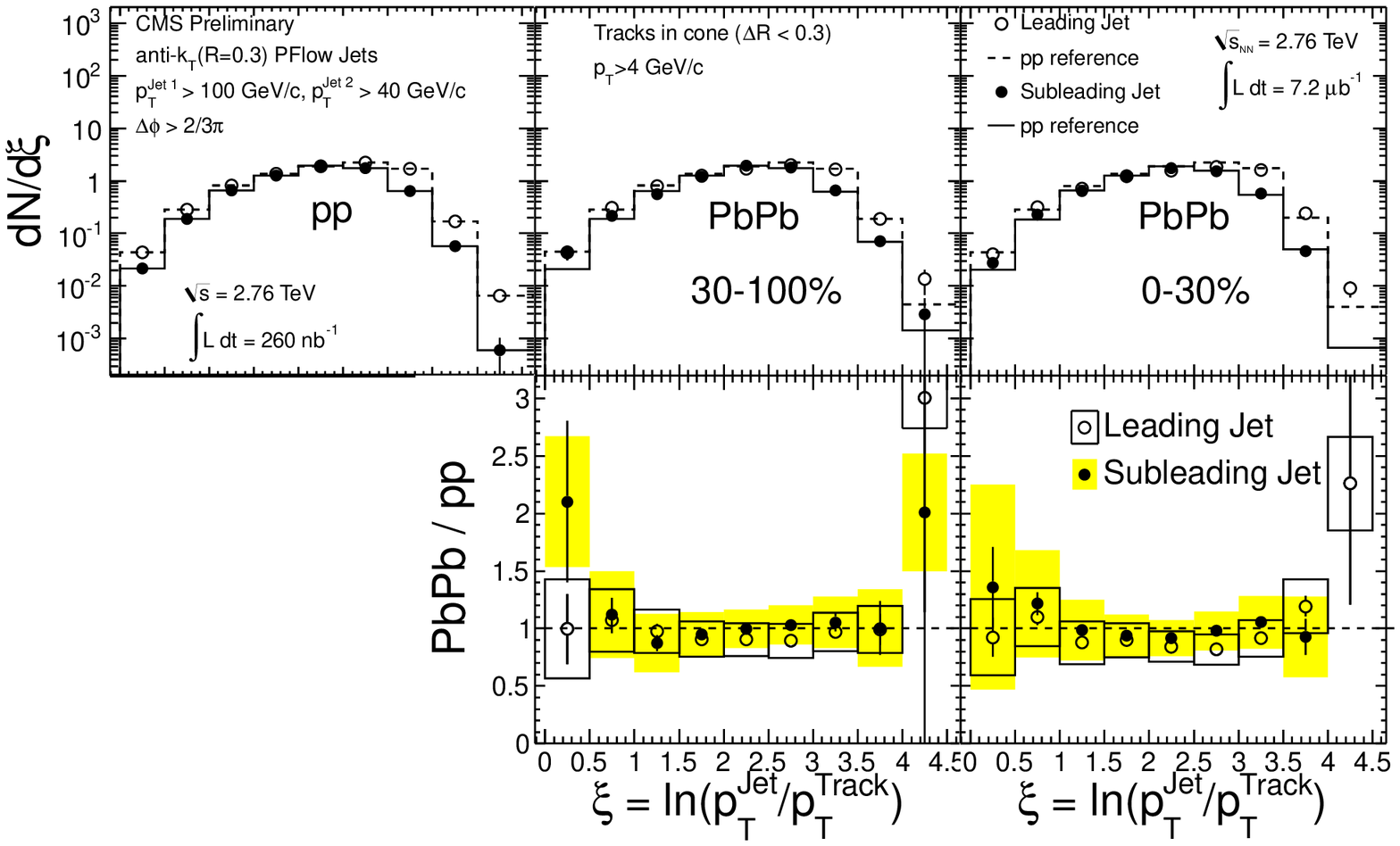}
\includegraphics[width=1.0\textwidth]{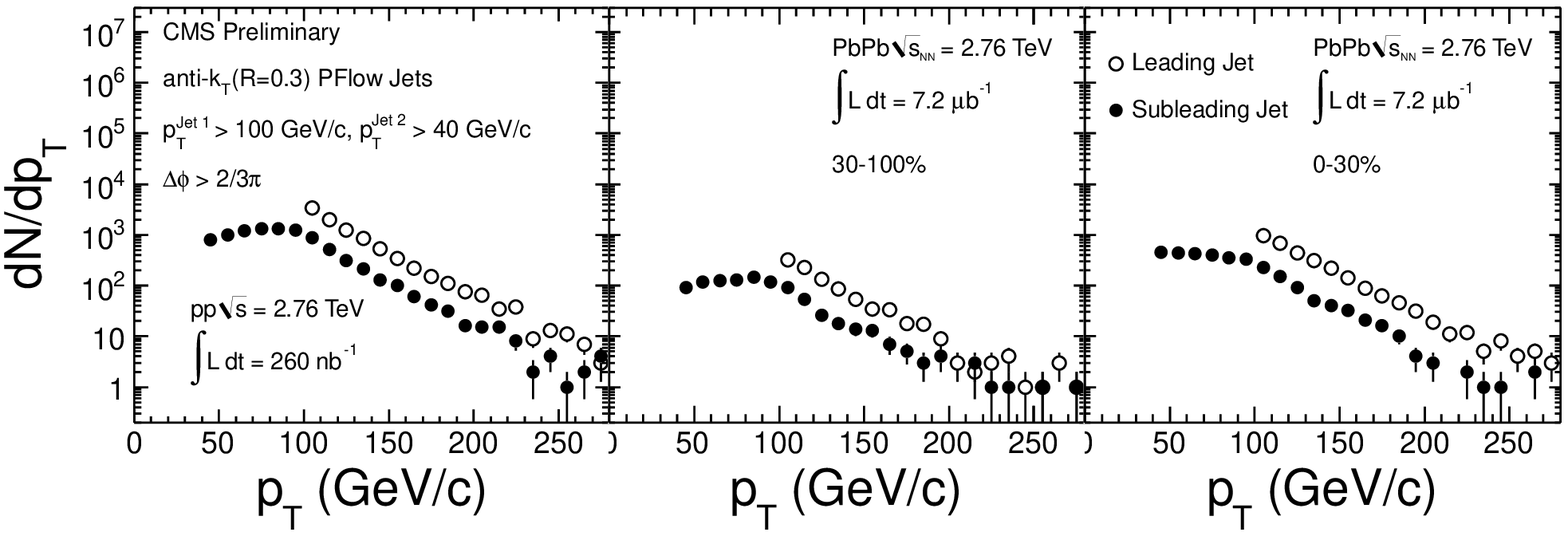}
\end{center}
\caption{Fragmentation functions in pp, peripheral PbPb, and central PbPb data, with the leading (open circles) and subleading (solid points) jets. The pp reference distributions are reweighted to match the jet $\pt$ distributions in PbPb and smeared to emulate background fluctuations in PbPb. The middle row shows the ratio of each PbPb fragmentation function to its pp reference.  Error bars represent statistical uncertainty, whereas the boxes represent the systematic uncertainty. The bottom row shows the jet $\pt$ distributions.}
\label{fig:ffLeadSublead_pp_ppDiv}
\end{figure}

\begin{figure}[b!]
\begin{center}
\includegraphics[width=1.0\textwidth]{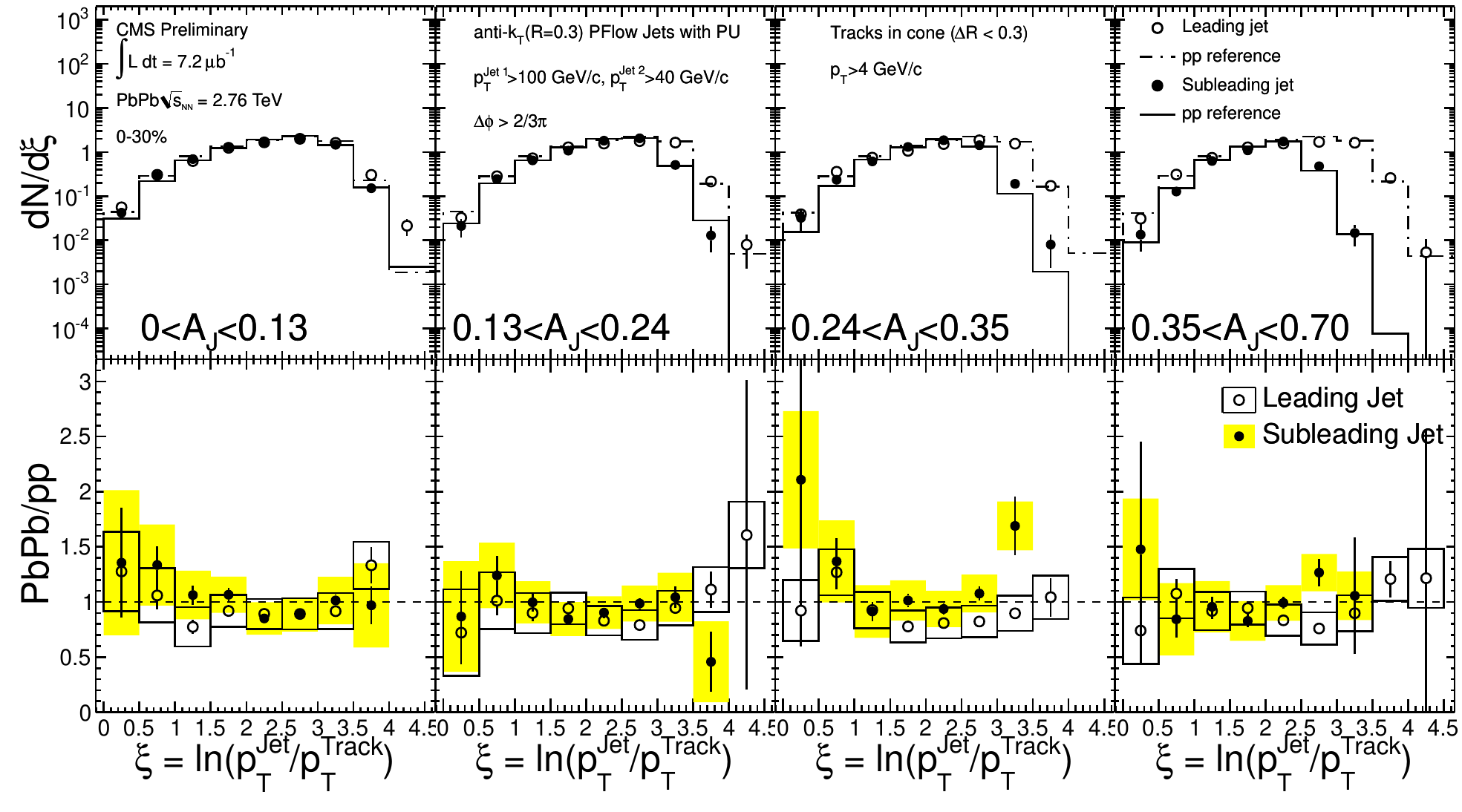}
\includegraphics[width=1.0\textwidth]{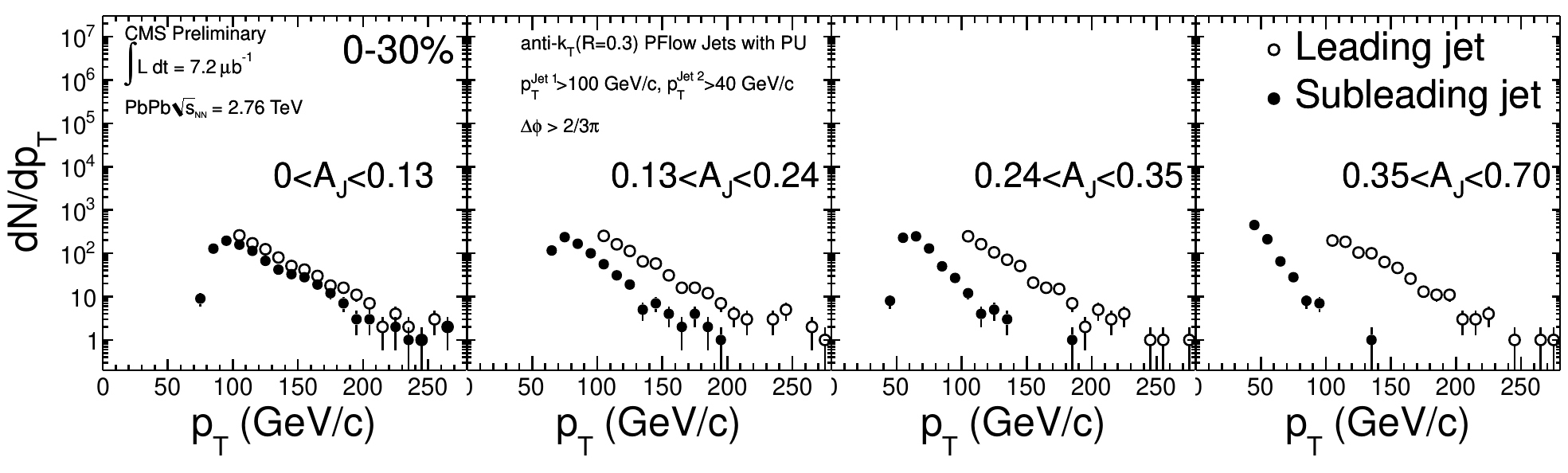}
\end{center}
\caption{Fragmentation functions in bins of $\AJ = (\pt^{1}-\pt^{2})/(\pt^{1}+\pt^{2})$, where $1$ and $2$ denote the leading and subleading jets in the event, reconstructed in central PbPb and pp reference for the leading
(open circles) and subleading (solid points) jets. The reference distributions are reweighted to match the jet $\pt$ distributions in PbPb and smeared to emulate background fluctuations in PbPb. The middle row shows the ratio of each fragmentation function to
its pp reference. The bottom row shows the jet $\pt$ distributions.}
\label{fig:ffLeadSublead_centralAJ_ppDiv}
\end{figure}


CMS~\cite{JINST} is a multi-purpose detector, with strong capabilities in tracking and calorimetry in a large geometric acceptance. The jets in this analysis were reconstructed using the CMS ``particle-flow" algorithm which attempts to identify all stable particles by combining information from all sub-detector systems~\cite{MattTalk:2011} and has the advantage of reduced sensitivity to the fragmentation pattern of jet, as compared to purely calorimetric jets. 

The anti-$k_{T}$ algorithm, as encoded in the
$FastJet$ framework, is used to combine the particle-flow candidates
into jets using a resolution parameter of R = 0.3~\cite{Cacciari:2008gp}. The small value of R helps to reduce background fluctuations.
The underlying background is subtracted using the ``Pile-up Subtraction Algorithm" described in~\cite{Kodolova:2007hd}.


This analysis uses heavy ion track reconstruction~\cite{SpectraPAS}. The tracking performance is studied within jets. Tracking efficiency tables are generated as a function of reconstructed jet $\pt$, reconstructed track $\pt$, track $\eta$, and collision centrality. The heavy ion track collection in general has a low fake rate ($<$ 5\%) with efficiency degrading from
$\sim$ 70\% at a few GeV/c to $\sim$ 50\% at 100 GeV/c.


The fragmentation functions in PbPb collisions are constructed and compared to those in pp collisions at $\roots = $ 2.76 TeV
For a direct comparison between pp and PbPb, that analysis of the pp data has to include the resolution deterioration due to the underlying event fluctuations seen in PbPb. For this purpose the pp jet \pt\ has been artificially smeared by the fluctuations observed in PbPb collisions taking into account the mean and RMS of the fluctuations as well as the correlation of the background level seen in back-to-back jet cones. In addition to the smearing, the pp reference is reweighted such that the final jet spectrum matches that of the PbPb events.
It is important to note that jet $\pt$ distributions of leading jets compared to subleading jets are different, which is also reflected in the respective fragmentation functions.


To study closely a potential effect of quenching on the fragmentation properties of partons, we divide the data sample into four  equally populated bins of dijet imbalance variable, $\AJ = (\pt^{1}-\pt^{2})/(\pt^{1}+\pt^{2})$, where $1$ and $2$ denote the leading and subleading jets in the event. The fragmentation functions are calculated for 0--30\% central events.
The fragmentation functions in central events are shown in figure~\ref{fig:ffLeadSublead_centralAJ_ppDiv}. The reference for these distributions is obtained by smearing the pp results and reweighting them such that the final leading or subleading jet spectrum matches that of the PbPb events in a given \AJ\ bin.



The hard component of fragmentation functions reconstructed in \PbPb\ collisions for different event centrality and dijet imbalance exhibit a universal behavior closely resembling the partons fragmenting in vacuum, as observed in pp collisions. The reconstructed fragmentation function in each bin of centrality and dijet imbalance are consistent with a jet of the same reconstructed \pt\ originating from a parton fragmenting without having traversed a nuclear medium, as seen in a comparison with pp collisions.

\section*{Bibliography}

\bibliographystyle{plain}
\bibliography{HIN_full}

\end{document}